\documentclass[aps,prl,twocolumn,groupedaddress,amsmath,amssymb,showpacs]{revtex4} 

\usepackage{graphicx}
\usepackage{dcolumn}
\usepackage{bm}

\begin{document}

\title{Topological Surface Transport in Epitaxial SnTe Thin Films Grown on Bi$_{2}$Te$_{3}$}

\author{A. A. Taskin}
\email{taskin@sanken.osaka-u.ac.jp}
\author{Fan Yang}
\author{Satoshi Sasaki}
\author{Kouji Segawa}
\author{Yoichi Ando}
\email{y_ando@sanken.osaka-u.ac.jp}

\affiliation{Institute of Scientific and Industrial Research,
Osaka University, Ibaraki, Osaka 567-0047, Japan}


\begin{abstract}

The topological crystalline insulator SnTe has been grown epitaxially on
a Bi$_{2}$Te$_{3}$ buffer layer by molecular beam epitaxy. In a
30-nm-thick SnTe film, $p$- and $n$-type carriers are found to coexist,
and Shubnikov--de Haas oscillation data suggest that the $n$-type
carriers are Dirac fermions residing on the SnTe (111) surface. This
transport observation of the topological surface state in a $p$-type
topological crystalline insulator became possible due to a downward band
bending on the free SnTe surface, which appears to be of intrinsic
origin.

\end{abstract}



\maketitle

The energy band inversion and time-reversal symmetry (TRS) are the main
ingredients for realizing a nontrivial topology in Z$_{2}$ topological
insulators (TIs) \cite{H-K, Moore, Q-SCZ, Ando}. Recently, the family of
TIs has been extended by the introduction of topological crystalline
insulators \cite{Fu2011,Hsieh2012} where the topology is protected by a
point-group symmetry of the crystal lattice rather than by TRS. The
first material predicted to be a TCI was SnTe \cite{Hsieh2012}, in which
the band inversion at an even number of time-reversal-invariant momenta
(TRIMs) leads to a trivial Z$_{2}$ topological invariant, but its mirror
symmetry gives rise to a nontrivial mirror Chern number $n_{M}$ = $-2$
to guarantee the existence of topologically protected gapless surface
states (SSs) on any surface containing a mirror plane. Angle-resolved
photoemission spectroscopy (ARPES) experiments have confirmed the existence of
Dirac-like SSs on the (001) surface of SnTe \cite{Tanaka2012} and
related compounds \cite{Dziawa2012,Xu2012}, generating a lot of interest
in TCIs \cite{QPI_SnTe, NanoWire_SnTe}. Naturally, an important next step
is to elucidate the topological SSs with transport experiments, as was
done for Z$_{2}$ TIs \cite{Taskin2009, Taskin2010, Qu2010, FisherNP,
BTS, BSTS, Morpurgo, Cui, KLWang, Ong2012}.

However, probing the SSs in SnTe by transport experiments is a
challenge, because of a high concentration of bulk holes (10$^{20}$ --
10$^{21}$ cm$^{-3}$) \cite{Nimtz}. Nevertheless, in thin films, an
enhanced surface-to-bulk ratio and a high surface mobility expected for
topologically-protected SSs \cite{DasSarma,PRL2012} might make it
possible to probe them in quantum oscillations. To obtain high-quality
thin films by molecular beam epitaxy (MBE)
\cite{MBE1,MBE2,MBE3,MBE4,MBE5,MBE6,MBE7,MBE8}, lattice matching of the
substrate is crucial. In this regard, while BaF$_{2}$ is the usual
choice of substrate for SnTe \cite{Ishida} with its $\sim$1.6\% lattice
matching, we noticed that rhombohedral Bi$_{2}$Te$_{3}$ may be a better
choice, at least for the (111) growth direction, with the lattice
matching of $\sim$1.5\%. Furthermore, the building block of
Bi$_{2}$Te$_{3}$ is a Te-Bi-Te-Bi-Te quintuple layer (QL) terminated
with a hexagonal Te plane, which naturally accommodates the Sn layer of
the SnTe in the (111) plane [see Fig. 1(d)]. 

Here, we show that high-quality SnTe thin films can indeed be grown by
MBE on Bi$_{2}$Te$_{3}$ and that they are actually suitable for probing
the topological SSs in transport experiments. Those films present
Shubnikov-de Haas (SdH) oscillations composed of two close frequencies,
whose dependence on the magnetic-field direction signifies that the
observed oscillations stem from two-dimensional (2D) Fermi surfaces
(FSs). Furthermore, the phase of the oscillations indicates that the 2D
carriers are Dirac electrons bearing the Berry phase of $\pi$.
Measurements of the $I$-$V$ characteristics across the
SnTe/Bi$_{2}$Te$_{3}$ interface and careful considerations of the
energy-band diagram in this heterostructure lead us to conclude that the
Dirac electrons reside on the top surface of SnTe.

The MBE growth was performed in an ultrahigh vacuum chamber with the
base pressure better than 5$\times$10$^{-8}$ Pa. Before deposition of
SnTe, a thin layer of high-quality Bi$_{2}$Te$_{3}$ was grown under
Te-rich conditions on sapphire substrates \cite{MBE-Te} with a two-step
deposition procedure similar to that used for Bi$_{2}$Se$_{3}$ films
\cite{MBE3,MBE5,AdvMater}. Both Bi (99.9999\%) and Te (99.9999\%) were
evaporated from standard Knudsen cells. The Te$_{2}$(Te$_{4}$)/Bi flux
ratio was kept at $\sim$20. The growth rate, which is determined by the
Bi flux, was kept at 0.3 nm/min. After growing $\sim$30 nm of the
Bi$_{2}$Te$_{3}$ layer, Sn (99.999\%) and Te were co-evaporated, keeping
the Te$_{2}$(Te$_{4}$)/Sn flux ratio at $\sim$40, substrate temperature
at 300$^{\circ}$C, and the growth rate at 0.4 nm/min. The resistivity
$\rho_{xx}$ and the Hall resistivity $\rho_{yx}$ of the films were
measured in a Hall-bar geometry by a standard six-probe method on
rectangular samples on which the contacts were made with silver paste or
indium near the perimeter. The magnetic field was swept between $\pm$14
T at fixed temperatures. 

\begin{figure}\includegraphics*[width=8.5cm]{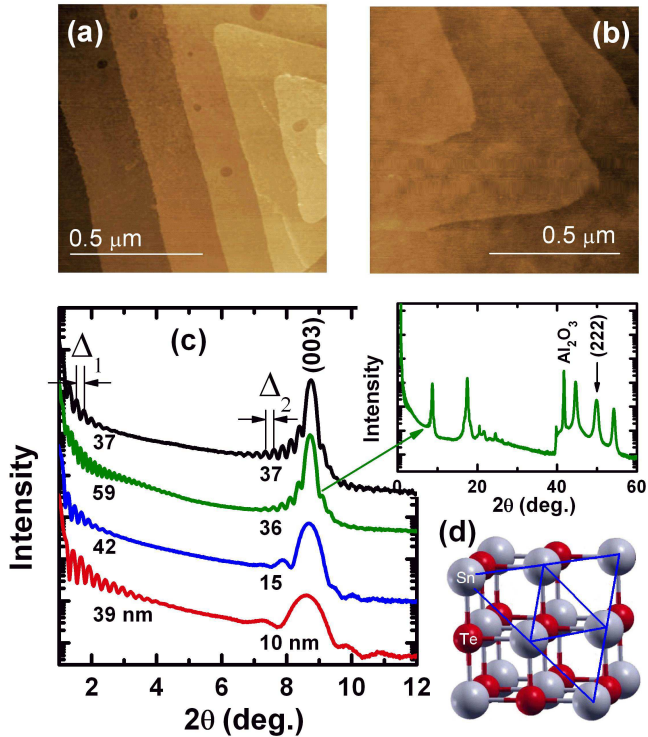}
\caption{(Color online) 
SnTe/Bi$_{2}$Te$_{3}$ heterostructure.
(a) AFM image of the Bi$_{2}$Te$_{3}$ layer showing
atomically flat terraces with 1-QL steps. 
(b) AFM image of the SnTe film grown on Bi$_{2}$Te$_{3}$ buffer layer.
The step height of the terraces is $\sim$0.4 nm.
(c) Low-angle XRD patterns of a series of SnTe films grown on
Bi$_{2}$Te$_{3}$ of different thickness. The total film thickness $d_t$
given by the distance $\Delta_{1}$ of Kiessig fringes at grazing angles
is shown to the left. The fringe distance $\Delta_{2}$ near the (003)
Bi$_{2}$Te$_{3}$ Bragg peak gives the thickness of the Bi$_{2}$Te$_{3}$
layer, $d_b$, which is shown near the peak. The SnTe layer thickness is
given by $d_t - d_b$. Inset shows a wide-angle XRD pattern.
(d) The rocksalt lattice of SnTe with its (111) plane marked by triangles.
}
\label{fig1}
\end{figure}

A critical ingredient for the epitaxial SnTe growth in the present
experiment is the high quality of the Bi$_{2}$Te$_{3}$ buffer layer.
Figure 1(a) shows an atomic force microscopy (AFM) image of a
40-nm-thick Bi$_{2}$Te$_{3}$ thin film grown on sapphire substrate.
Large equilateral triangles with atomically flat terraces, which have a
height of exactly 1 QL, can be easily recognized. An
AFM image of a 30-nm-thick SnTe film grown on top of such
Bi$_{2}$Te$_{3}$ buffer layer is shown in Fig. 1(b). Triangles are still
clearly seen on the surface, giving evidence for an epitaxial growth.
The height of the terraces is $\sim$0.4 nm, which agrees with the
periodicity of the rock-salt lattice along the (111) direction [Fig.
1(d)]. (An image for a larger area with clear triangular morphology is
shown in the Supplemental Material (SM) \cite{SM}).

The high structural quality of both Bi$_{2}$Te$_{3}$ and SnTe films as
well as the very smooth nature of the interface between them can be
judged from the Kiessig fringes \cite{SM,Pietsch} in the x-ray
diffraction (XRD) measurements [Fig. 1(c); see also SM for more details]. 
The inset of Fig. 1(c) shows the XRD pattern for a wider angle range, 
in which SnTe only yields $(2n,2n,2n)$ Bragg peaks to confirm the (111) 
growth direction.

\begin{figure}\includegraphics*[width=8.5cm]{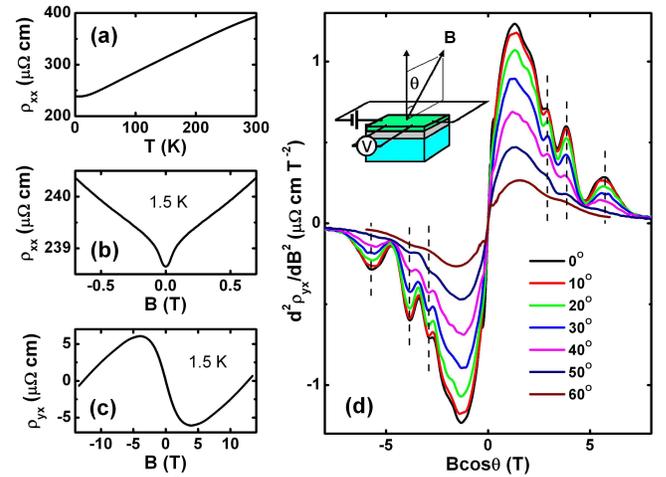}
\caption{(Color online) 
Transport properties of a 30-nm-thick SnTe film grown on a 36-nm 
Bi$_{2}$Te$_{3}$ buffer layer.
(a) Temperature dependence of $\rho_{xx}$. 
(b) Low-field $\rho_{xx}(B)$ measured at 1.5 K.
(c) $\rho_{yx}(B)$ measured at 1.5 K.
(d) $d^{2}\rho_{yx}/dB^{2}$ at various angles are plotted vs $B \cos \theta$; 
inset shows the measurement geometry. 
Dashed lines mark the maxima in the oscillations.
}
\label{fig2}
\end{figure}

Figure 2(a) shows the temperature dependence of the resistivity,
$\rho_{xx}(T)$, in a 30-nm-thick SnTe film grown on 36-nm-thick
Bi$_{2}$Te$_{3}$. There is no discernible kink in the data, suggesting
that the structural phase transition observed in bulk SnTe
\cite{Nimtz,Kobayashi,Katayama} is absent in our thin films and that the
mirror symmetry is kept intact down to low temperature \cite{SM}. In the
magnetotransport properties, a downward cusp observed in $\rho_{xx}(B)$
at very low fields [Fig. 2(b)] is a reflection of the weak
antilocalization behavior which is expected for topological materials
\cite{Chen,WAL1,WAL2}. We also observe a coexistence of $n$- and $p$-type
carriers in the sample which is evident from a sign change of the slope
in $\rho_{yx}(B)$ [Fig. 2(c)]. Importantly, we found that both
$\rho_{yx}(B)$ and $\rho_{xx}(B)$ present SdH oscillations at high
magnetic fields. To remove a large background and make the oscillations
more visible, we employed second derivatives. Figure 2(d) shows
$d^{2}\rho_{yx}/dB^{2}$ measured in tilted magnetic fields at 1.5 K and
plotted as a function of $B \cos \theta$, where $\theta$ is the angle of
the magnetic field from the surface normal. Since the maxima in the
oscillations (marked by vertical dashed lines) appear at the same $B
\cos \theta$ upon changing $\theta$, the observed SdH oscillations
clearly have a two-dimensional (2D) character. Note also that in our
experiments, the SdH oscillations were not seen at tilting angles close
to 90$^{\circ}$, giving evidence against a three-dimensional (3D) FS as
the origin of oscillations.

An important question is which of the $n$- or $p$-type carriers are
responsible for the oscillations, and this can be answered in the following
Landau-level (LL) index analysis. To properly construct the LL index
plot, we use conductance $G_{xx}$ and Hall conductance $G_{xy}$ rather
than $\rho_{xx}$ and $\rho_{yx}$ \cite{Ando}. Figure 3(a) shows the
plots of $d^{2}G_{xx}/dB^{2}$ and $d^{2}G_{xy}/dB^{2}$ vs 1/$B$. The
Fourier transform of $d^{2}G_{xy}/dB^{2}$ is shown in the inset of Fig
3(a). Its main feature is a broadened peak with a shoulder, which can be
well fitted with two Gaussians centered at frequencies of 10.6 and 14 T.
The coexistence of two branches of oscillations is actually anticipated
from weak beating patterns in the data. The two frequencies $F_{1}$ =
10.6 T and $F_{2}$ = 14 T correspond to orbits on the FSs with radii of
$k_{F}$ = 1.8 $\times$ 10$^{6}$ cm$^{-1}$ and 2.1 $\times$ 10$^{6}$
cm$^{-1}$, respectively. The corresponding 2D carrier densities $n_{s}$
are 2.6 $\times$ 10$^{11}$ cm$^{-2}$ and 3.4 $\times$ 10$^{11}$
cm$^{-2}$ for each spin eigenvalue. Since the amplitude of the
lower frequency oscillations is much larger than the amplitude of the
higher frequency branch [see inset of Fig. 3(a)], the main contribution
to the SdH oscillations is coming from the lower frequency branch; in
such a case, the LL index plot constructed from weakly beating
oscillations can still yield the correct phase factor for the
lower frequency branch with reasonable accuracy (see SM for details).
The constructed LL index plot [Fig. 3(b)] crosses the $N$-index axis at
0.55, which gives evidence for the Berry phase of $\pi$
\cite{TA_BerryPhase,MS_BerryPhase, WM_BerryPhase}. Also, the relative
phase in the oscillations of $d^{2}G_{xx}/dB^{2}$ and
$d^{2}G_{xy}/dB^{2}$ indicates that the carriers must be $n$-type (see SM
for details). Therefore, the observed SdH oscillations can be concluded
to be due to $n$-type 2D Dirac fermions bearing the $\pi$ Berry phase.
Note that, even though the Bi$_{2}$Te$_{3}$ layer contains a lot of
$n$-type carriers (see SM), such carriers cannot be the source of the
SdH oscillations, because the observed frequencies are an order of
magnitude too low to represent the bulk FS of Bi$_{2}$Te$_{3}$.

\begin{figure}\includegraphics*[width=8.5cm]{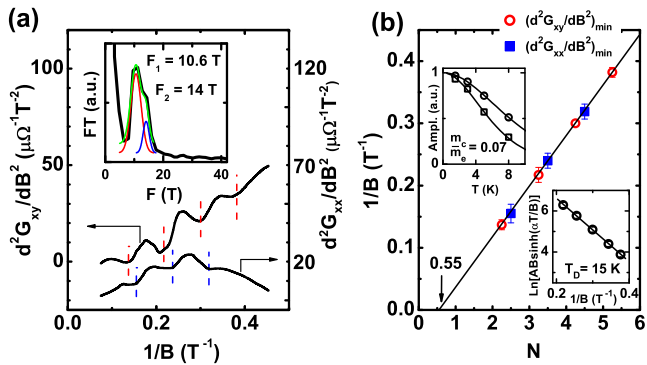}
\caption{(Color online) 
SdH oscillations.
(a) $d^{2}G_{xx}/dB^{2}$ and $d^{2}G_{xy}/dB^{2}$ vs 1/$B$ measured at
$T$ = 1.5 K and $\theta$ = 0$^{\circ}$; inset shows the Fourier
transform of the $d^{2}G_{xy}/dB^{2}$ oscillations revealing two close
frequencies, 10.6 and 14 T (the upturn below $\sim$5 T is
related to the background which slowly changes with $B$). 
(b) LL index plot constructed from the minima in the oscillations of
$d^{2}G_{xx}/dB^{2}$ and $d^{2}G_{xy}/dB^{2}$. A half-integer index
$N+\frac{1}{2}$ is assigned to a minimum in $d^{2}G_{xx}/dB^{2}$. The
index assignment for a minimum in $d^{2}G_{xy}/dB^{2}$ depends on the
sign of the carriers: The index $N+\frac{1}{4}$ for electrons is
consistent with the indices from $G_{xx}$, meaning that the SdH
oscillations are produced by electrons. The solid line is a linear
fitting to the data; its intercept of 0.55 on the $N$-index axis
indicates the $\pi$ Berry phase. Upper inset shows $T$ dependencies of
the SdH amplitudes measured at 2.67 T (squares) and 3.85 T (circles),
both yielding $m_{c}$ = 0.07$m_{0}$. Lower inset shows the Dingle plot
for the data at 1.5 K, giving $T_{D}$ = 15 K.
}
\label{fig3}
\end{figure}

The temperature dependence of the SdH amplitude [upper inset of Fig.
3(b)] gives the cyclotron mass $m_{c}$ = 0.07$m_{0}$ ($m_0$ is the free
electron mass) \cite{Shoenberg}. This value should mainly reflect the
lower frequency branch of oscillations ($F_{1}$ = 10.6 T) due to its
dominance in the data, and we conclude that the upper limit of the Fermi
velocity $v_{F}$ (= $\hbar k_{F}/m_{c}$) of the dominant surface
carriers is about 3 $\times$ 10$^{7}$ cm/s. The Dingle analysis [lower
inset of Fig. 3(b)] yields the Dingle temperature $T_{D}$ of 15 K, from
which the mean-free path of Dirac electrons $l^{\rm SdH}$ = 24 nm and
their mobility $\mu_{s}^{\rm SdH}$ = 2000 cm$^2$/Vs are calculated
\cite{Ando}. Such a mobility is typical for best-quality SnTe films
\cite{Ishida}.

\begin{figure}\includegraphics*[width=8.5cm]{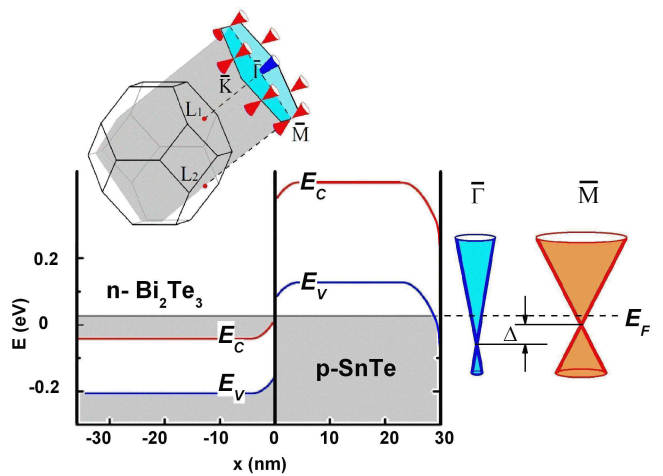}
\caption{(Color online) 
Energy-band diagram of $n$-Bi$_{2}$Te$_{3}$/$p$-SnTe heterojunction. The
broken-gap band lineup is concluded from the $I$-$V$ characteristics of the
heterojunction interface (see SM for details).
Downward band-bending on the free surface of SnTe gives rise to $n$-type
doping of surface Dirac cones shown schematically on the right. The
upper inset shows the bulk Brillouin zone of SnTe and its projection
along the (111) direction to the surface Brillouin zone, which hosts two
kinds of Dirac cones at $\bar{\Gamma}$ and $\bar{M}$. The shaded plane
is one of the three mirror planes \{110\}.
}
\label{fig4}
\end{figure}

Now we discuss the origin of the observed $n$-type Dirac fermions. Both
Bi$_{2}$Te$_{3}$ and SnTe have topological surface states, and it is
useful to consider the energy-band diagram of the heterojunction (shown
in Fig. 4) formed by degenerate $p^+$-SnTe grown on the degenerate
$n^+$-Bi$_{2}$Te$_{3}$. The lineup of the conduction and valence bands
at the interface of two semiconductors is of fundamental importance for
understanding the properties of the heterojunction. Essentially, there
are three possibilities: straddling, staggered, and broken-gap band
lineups \cite{KroemerNL} (see SM for details). The vast majority of
heterojunctions have a straddling lineup with conduction- and
valence-band offsets of opposite sign; in this case, when the two sides
are doped with opposite types of carriers, an insulating barrier layer
will be formed at the interface of such a $p-n$ junction. The same holds
true for the case of a staggered lineup, in which conduction- and
valence-band offsets have the same sign with a finite overlap of the
gaps. The situation is different for the most exotic broken-gap lineup,
in which the bottom of the conduction band of one semiconductor goes
below the top of the valence band of the other semiconductor as has been
shown for InAs/GaSb heterostructures \cite{Sakaki1977}. In this case,
the system can behave as a semimetal without forming any barrier at the
interface of a $p-n$ junction.

To determine which of the possible lineups is realized in our system, we
measured $I$-$V$ curves across the interface in a sample where a part of
the SnTe film has been etched away to make direct electrical contacts to
both Bi$_2$Te$_3$ and SnTe layers (see SM). We found the $I$-$V$
characteristics to show Ohmic behavior, which led us to conclude that
the SnTe/Bi$_2$Te$_3$ heterojunction most likely has the broken-gap
lineup. In such a case, the Fermi level at the interface may lie above
the bottom of the conduction band of Bi$_2$Te$_3$ and below the top of
the valence band of SnTe. Hence, while some exotic 2D state may be
formed at the Bi$_2$Te$_3$/SnTe interface \cite{Soriano, Eremeev,
Koleini}, such a state is not accessible due to the position of the Fermi
level and it is unlikely that the 2D SdH oscillations come from this
interface.

Another interface between Bi$_{2}$Te$_{3}$ and sapphire is also an
unlikely place for $n$-type Dirac fermions to reside on, because the
Dirac point of the SS in Bi$_{2}$Te$_{3}$ is situated below the top of
its valence band \cite{BT_ARPES}. This means that, in order for the SdH
oscillations with frequencies of only 10 -- 14 T to be observed, a very
large upward band bending sufficient for creating an inversion layer
would be required at the interface with sapphire. This is very unlikely
and, in fact, we have never observed such low-frequency SdH oscillations
in Bi$_{2}$Te$_{3}$ films grown on sapphire.
 
Therefore, the only viable possibility is that the top SnTe surface has 
a sufficient downward band bending (Fig. 4) to host $n$-type Dirac fermions. 
Interestingly, such a band bending is naturally expected in materials with 
partially ionic bonding. For Sn$^{2+}$Te$^{2-}$ films grown in the [111] 
direction, the stacking sequence of atomic planes is 
Sn$^{2+}$-Te$^{2-}$-$\cdot\cdot\cdot$, which brings about a dipole moment 
and leads to a diverging electrostatic energy (see SM). This situation is 
known as the polar catastrophe  \cite{Weiss} and cannot be realized in real 
materials; what actually happens is a partial charge compensation on the 
top and bottom surfaces to avoid the accumulation of electrostatic potential. 
In our system, the first atomic plane of the SnTe layer at the interface 
should be composed of Sn$^{2+}$ and some of its positive charge is naturally 
compensated by $n$-type carriers of the Bi$_{2}$Te$_{3}$ layer. 
On the free surface side, the termination is either with Te$^{2-}$ or 
Sn$^{2+}$ planes; since the SnTe layer begins with Sn$^{2+}$, 
the termination with Te$^{2-}$ costs more electrostatic energy 
and Sn$^{2+}$ termination is preferable (see SM). The resulting charge 
compensation leads to a downward band-bending at the Sn$^{2+}$-terminated 
free surface as shown in Fig. 4. This offers a natural explanation of the 
observed $n$-type carriers at the free SnTe surface. 
Note that a strong downward band banding is also observed in ARPES 
experiments \cite{Tanaka_SnTe111} when $p$-type SnTe single crystals are 
cleaved along the [111] direction in vacuum.

The above picture allows us to consistently understand the measured
transport data. On the (111) plane of SnTe which is a TCI, there are
four Dirac cones centered at four TRIMs in the surface Brillouin zone
(BZ) \cite{Tanaka_SnTe111,JunweiLiu_PRB}: one at $\bar{\Gamma}$ and
three at $\bar{M}$ points which are projections of the four $L$ points
in the 3D BZ along the [111] direction as schematically shown in the
inset of Fig. 4. The surface band calculations give different results
for Te and Sn terminations \cite{JunweiLiu_PRB,Polish_PRB}. For
Te-terminated (111) surface, all Dirac points (DPs) touch the bottom of
the conduction band, and it is impossible to realize $n$-type Dirac
fermions irrespective of the position of the Fermi level. For the
Sn-terminated (111) surface, on the other hand, the DPs are closer to
the top of the valance band and Dirac electrons can be probed in
transport experiments. Interestingly, epitaxially grown (111)-oriented
films of a similar material, Pb$_{1-x}$Sn$_{x}$Se, were found to be
preferentially terminated with Pb/Sn \cite{Polley_PbSnSe}.

The observed two frequencies in the SdH oscillations are consistent with
the existence of two types of Dirac cones on the free surface of SnTe
reported in ARPES experiments \cite{Tanaka_SnTe111}: the stronger,
lower frequency branch is coming from electrons occupying the three
Dirac cones at the $\bar{M}$ points, while the weaker, higher frequency
branch is coming from electrons occupying the sole Dirac cone at the
$\bar{\Gamma}$ point. Importantly, the ARPES data
show \cite{Tanaka_SnTe111} that the Dirac point at $\bar{\Gamma}$ is
lower in energy than that at $\bar{M}$, resulting in a higher Fermi
energy for the Dirac cone at $\bar{\Gamma}$ (see Fig. 4) with a
difference $\Delta$ of $\sim$170 meV. It is worth noting that if
oscillations were coming from the Bi$_{2}$Te$_{3}$ surface, there would
be only one frequency. The same is true for the SnTe/Bi$_{2}$Te$_{3}$
interface, where the two Dirac cones at $\bar{\Gamma}$ originating from
SnTe and Bi$_{2}$Te$_{3}$ should annihilate due to their opposite
helicities \cite{Germ_Interface}. The $v_{F}$ of about 3 $\times$
10$^{7}$ cm/s obtained from our data may be attribute to the averaged
$v_{F}$ of highly anisotropic DPs at $\bar{M}$ \cite{Polish_PRB}. From
$F_{1}$ = 10.6 T ($k_{F}$ = 1.8 $\times$ 10$^{6}$ cm$^{-1}$), the
position of the Fermi level above the DPs at $\bar{M}$ is estimated to
be about 40 meV. For the DP at $\bar{\Gamma}$, according to the ARPES
data \cite{Tanaka_SnTe111}, the Fermi velocity is much larger, $v_{F}$ =
1.3 $\times$ 10$^{7}$ cm/s, and, for $F_{2}$ = 14 T ($k_{F}$ = 2.1
$\times$ 10$^{6}$ cm$^{-1}$), the position of the Fermi level would be
about 180 meV above the DP. The energy difference of $\sim$140 meV
between the two types of Dirac cones obtained in our transport
experiments is close to the ARPES result of $\Delta \sim$170 meV, giving
confidence that the observed 2D electrons indeed reside on the free
surface of SnTe.

Finally, we mention that the observed SdH oscillations are prone to
aging; namely, their amplitude was greatly reduced when we re-measured
the sample after keeping it in nitrogen atmosphere for six months. This
also supports the conclusion that the 2D oscillations are most likely
coming from the free surface of SnTe. All in all, the present results
demonstrate that the surface Dirac electrons residing on the (111)
surface of SnTe can be accessed by transport measurements of
high-quality films grown on a Bi$_{2}$Te$_{3}$ buffer layer. These 
thin-film samples open new opportunities for experimentally exploring
the physics of TCIs as well as for fabricating novel devices based on
the unique nature of TCIs \cite{Liu_MBE, Fang_MBE}.

\begin{acknowledgments} 

We thank J. Liu and L. Fu for helpful discussions, and M. Kishi for
technical assistance with micro-fabrication of samples for $I$-$V$
measurements. This work was supported by JSPS (KAKENHI 24740237,
24540320, 25400328, and 25220708), MEXT (Innovative Area ``Topological
Quantum Phenomena" KAKENHI), and AFOSR (AOARD 124038).

\end{acknowledgments}

\clearpage
\onecolumngrid

\renewcommand{\thefigure}{S\arabic{figure}} 

\setcounter{figure}{0}

\renewcommand{\thesection}{S\arabic{section}.} 

\begin{flushleft} 
{\Large {\bf Supplemental Material}}
\end{flushleft} 

\vspace{2mm}

\begin{flushleft} 
{\bf S1. Resistivity behavior of Bi$_{2}$Te$_{3}$ buffer layer}
\end{flushleft} 

Our Bi$_{2}$Te$_{3}$ films grown on sapphire substrates with MBE are
always degenerately $n$-type doped. Figure S1 shows typical data of the
temperature dependence of the resistivity $\rho_{xx}$. The bulk electron
density $n_b$ of this films is known from the Hall effect to be 4
$\times$ 10$^{19}$ cm$^{-3}$, from which one obtains the bulk mobility
of about 400 and 100 cm$^2$/Vs at 1.5 and 300 K, respectively.

\begin{figure}[b]
\begin{center}
\includegraphics[height=7cm]{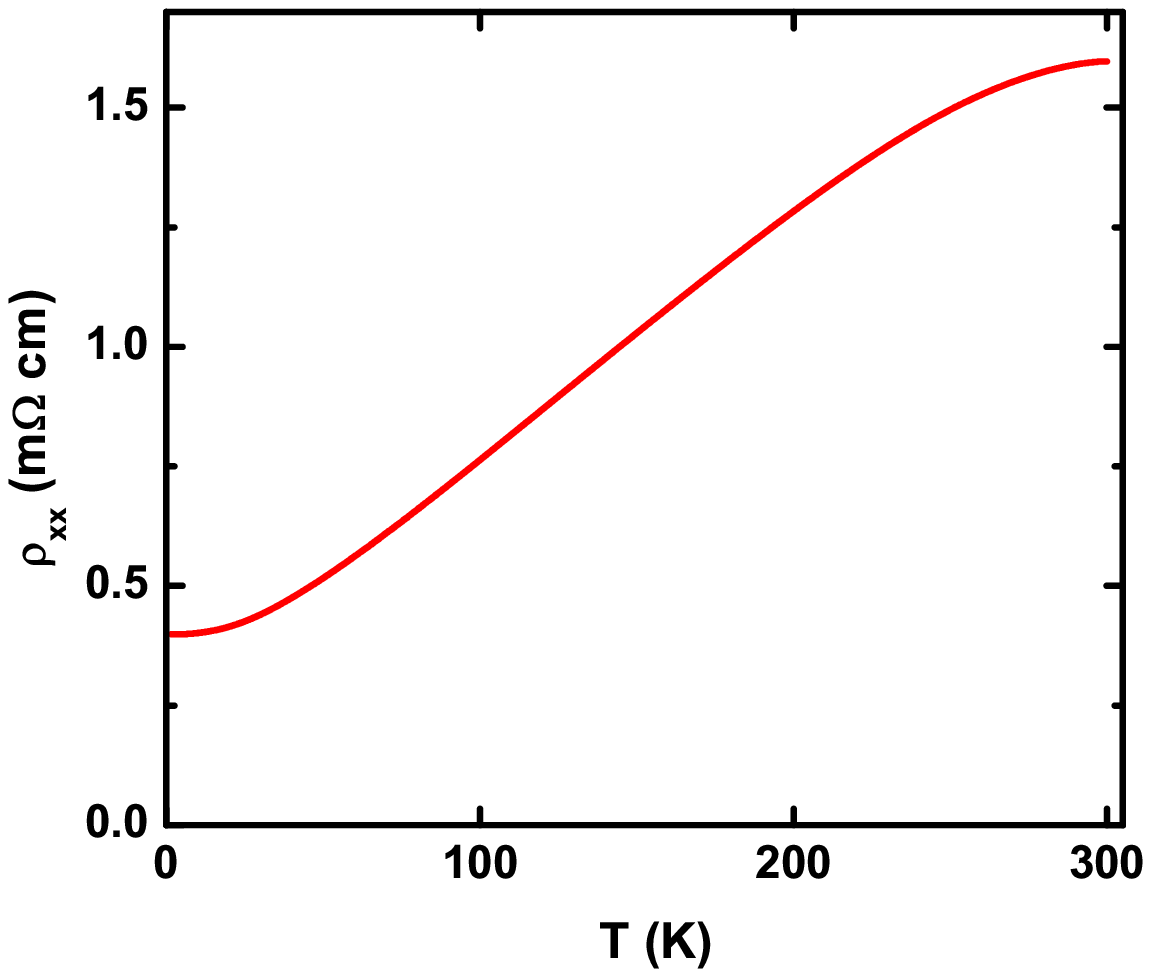}
\caption{
Temperature dependence of $\rho_{xx}$ of a 38-nm-thick
Bi$_{2}$Te$_{3}$ film without SnTe deposition.
} 
\end{center}
\end{figure}

\begin{flushleft} 
{\bf S2. Kiessig fringes}
\end{flushleft} 

Low-angle XRD patterns shown in Fig. 1{\bf c} in the main text demonstrate
clear Kiessig oscillations. Generally, such low-angle intensity
oscillations are the result of x-ray interference between two interfaces
of a thin film, indicating its homogeneity (uniform structure) and
smoothness of the interfaces \cite{Pietsch,AdvMater}. The position of
the $m$-th maximum $\theta_{m}$ in Kiessig fringes follows \cite{Pietsch}
\begin{equation}
\theta_{m}^{2}=\alpha_{c}^2+m^{2} \left(\frac{\lambda}{2t} \right) ^2,
\end{equation}
where $\alpha_{c}$ is the critical angle for the total external reflection, 
$\lambda$ is the x-ray wavelength, and $t$ is the thickness of a film.
Since $\alpha_{c} \approx$ 0 for x-rays, the film thickness 
can be accurately estimated as 
\begin{equation}
t \approx \lambda/2 \Delta,
\end{equation}
where $\Delta$ is the period of Kiessig oscillations (i.e. fringe distance) 
in radian unit. 

\begin{figure}
\begin{center}
\includegraphics[height=5.5cm]{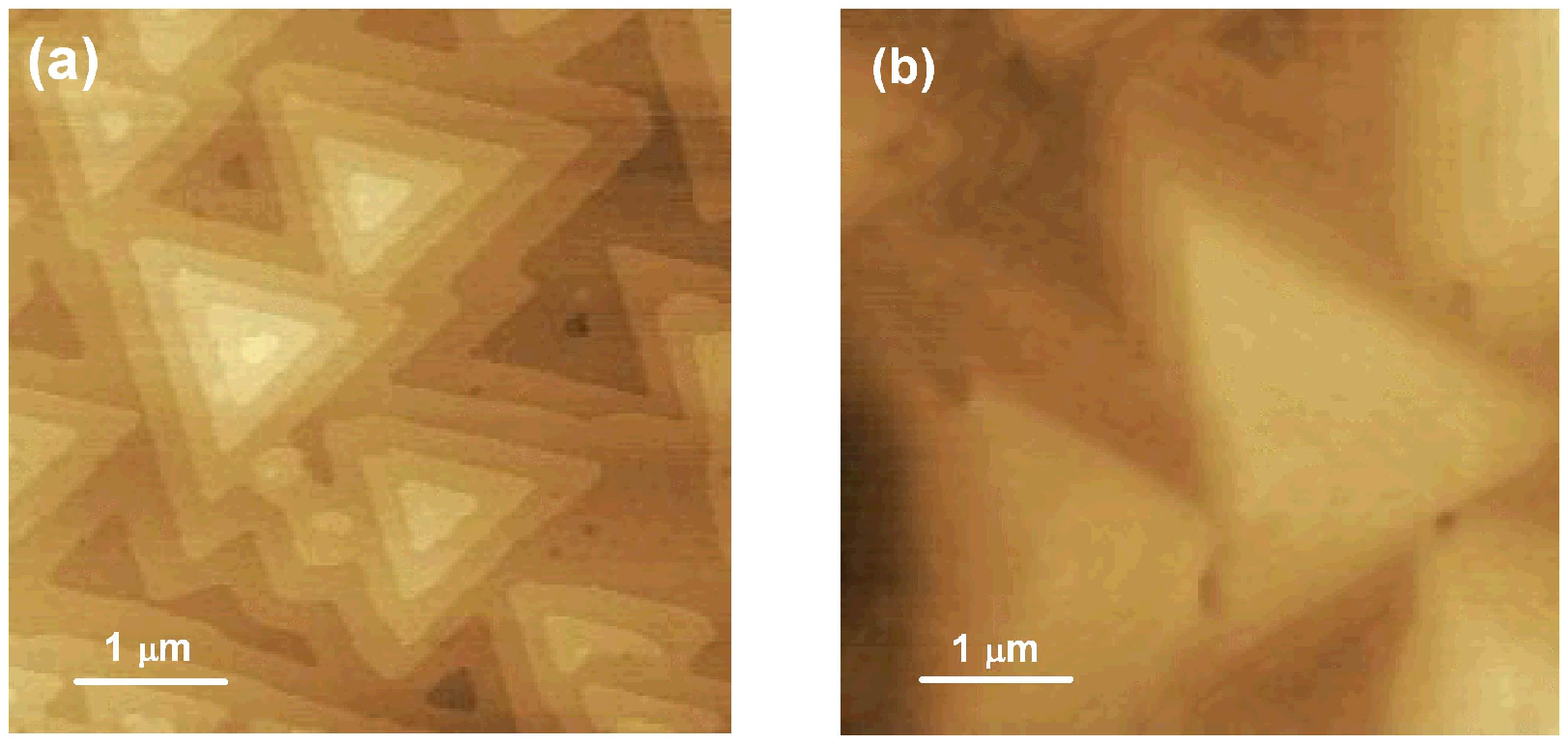}
\caption{
AFM images of ({\bf a}) 40-nm-thick Bi$_{2}$Te$_{3}$ film and ({\bf b}) 40-nm-thick 
SnTe film on top of Bi$_{2}$Te$_{3}$ over a large area of about 5 $\times$ 5
$\mu$m$^2$.
} 
\end{center}
\end{figure}

Our samples consist of sapphire (Al$_2$O$_3$) substrate,
Bi$_{2}$Te$_{3}$ buffer layer, and SnTe layer; typical morphologies of
Bi$_{2}$Te$_{3}$ and SnTe films over a large area of about 5 $\times$ 5
$\mu$m$^2$ are shown in Fig. S2. The fringe distance observed at grazing
incidence angles reflects the total thickness $d_t$ of the film, namely,
the thickness of Bi$_{2}$Te$_{3}$ layer plus SnTe layer, and indeed, it
exactly matches the thicknesses measured by AFM near a sharp edge.
Interestingly, in our samples, Kiessig fringes are seen not only at
grazing angles, but also near the (003) Bragg peak of Bi$_{2}$Te$_{3}$
layer. They have larger fringe distances which are the results of the
interference within the Bi$_{2}$Te$_{3}$ buffer layer; hence, they
testify to the high quality of both Al$_{2}$O$_{3}$/Bi$_{2}$Te$_{3}$ and
Bi$_{2}$Te$_{3}$/SnTe interfaces and allow us to estimate the thickness
$d_b$ of the Bi$_{2}$Te$_{3}$ layer alone. [The top curve in Fig. 1{\bf c}
shows the data for Bi$_{2}$Te$_{3}$ film without deposition of SnTe, and
hence the fringe distances at grazing angles and near the (003) peak are
the same.] Since both $d_t$ and $d_b$ are known from those Kiessig
fringes, the SnTe layer thickness is accurately estimated by calculating
$d_t - d_b$.


\begin{flushleft} 
{\bf S3. Ferroelectric transition in SnTe}
\end{flushleft} 

SnTe is known to undergo a ferroelectric transition at low temperature,
which is associated with a structural phase transition from cubic to
rhombohedral \cite{Nimtz,Kobayashi,Katayama}. In bulk single crystals,
this ferroelectric transition manifests itself in the temperature
dependence of $\rho_{xx}$ as a kink \cite{Kobayashi}. In our thin films,
however, the resistivity data do not present any feature ascribable to
the ferroelectric transition, which suggests that the structural transition
is suppressed by the epitaxial strain. The absence of the ferroelectric
transition in strained films implies that the mirror symmetry is kept
intact down to low temperature. Actually, even if the ferroelectric
transition happens, it will {\it not} break the mirror symmetry on the
surface of our films, because the structural distortion occurs along the
[111] direction which is the same as the growth direction.

\begin{flushleft} 
{\bf S4. Assignment of Landau-level indices}
\end{flushleft} 

As discussed in detail in Ref. \cite{Ando}, for the assignment of
Landau-level (LL) indices in the data of Shubnikov-de Haas (SdH)
oscillations, the best practice is to start with the principle that an
integer index $N$ be assigned to a minimum in the conductance $G_{xx}$,
and it follows that $N+\frac{1}{2}$ be assigned to a minimum in
$d^{2}G_{xx}/dB^{2}$. The index assignment for the Hall conductance
$G_{xy}$ depends on the sign of the carriers, and the correct assignment
is most easily understood by considering the integer quantum Hall
effect: For electrons, $G_{xy}$ {\it increases} with $B$ between the
Hall plateaus and hence $dG_{xy}/dB$ shows a {\it maximum} at
$N+\frac{1}{2}$, whereas for holes $G_{xy}$ {\it decreases} between
plateaus to cause $dG_{xy}/dB$ to show a {\it minimum} at
$N+\frac{1}{2}$. Therefore, when the index assignment is to be done to a
minimum in $d^{2}G_{xy}/dB^{2}$, the index should be $N+\frac{1}{4}$ if
the carriers are electrons, while it should be $N+\frac{3}{4}$ if they
are holes. (Note that the LL index {\it decreases} with increasing $B$,
which is the reason why a minimum in $d^{2}G_{xy}/dB^{2}$ should be at
$N+\frac{1}{4}$ when a maximum in $dG_{xy}/dB$ is at $N+\frac{1}{2}$.)

In the case of the data shown in Fig. 3 of the main text, it turns out
that indices for electrons, $N+\frac{1}{4}$, should be assigned to 
the minima in $d^{2}G_{xy}/dB^{2}$ to make consistency with
the indices from $d^{2}G_{xx}/dB^{2}$, meaning that the SdH oscillations are
produced by electrons.


\begin{flushleft} 
{\bf S5. Landau-level index plot for weakly beating SdH oscillations}
\end{flushleft} 

To understand the behavior of weakly beating SdH oscillations composed
of two similar frequencies, let us consider a sum of two oscillating
parts in the conductivity, $A_1\cos\left(2\pi \frac{F}{B}\right)$ and
$A_2\cos\left(2\pi \frac{F+\delta}{B} + \alpha \right)$. In our data,
$F$ = 10.6 T and $\delta$ = 3.4 T, and the amplitude $A_1$ is
about three times larger than $A_2$. The factor $\alpha$ is a possible
phase shift between the two branches of oscillations. We simplify the 
calculations by setting the phase of the main branch to be 0. One can 
see that the sum of the two is also a cosine function, namely,
\begin{equation}
A_1\cos\left(2\pi \frac{F}{B}\right) + A_2\cos\left(2\pi \frac{F+\delta}{B}+ \alpha \right) 
= A\cos\left(2\pi \frac{F}{B} + \phi\right),
\end{equation}
where the amplitude $A$ and the phase $\phi$ are given by
\begin{align}
A &= \sqrt{A_1^{2}+A_2^{2}+2A_1 A_2\cos(2\pi \frac{\delta}{B}+ \alpha )}\,,\\
\phi &= \arctan \left[\frac{A_2\sin(2\pi \frac{\delta}{B}+ \alpha)}
{A_1 + A_2 \cos(2\pi \frac{\delta}{B}+ \alpha)} \right] .
\label{eq:phase}
\end{align}
The total amplitude $A$ is not constant, but changes periodically
between $A_1-A_2$ and $A_1+A_2$, which is the reason for beating. The
new phase $\phi$ is also an oscillating function. From Eq.
(\ref{eq:phase}), it is easy to see that $|\phi|$ cannot exceed the
value of $\arctan\frac{A_2}{A_1-A_2}$, which is less than 0.15$\pi$ for
$A_1:A_2 = 3:1$. This means that the phase of the weakly beating
oscillations essentially follow the phase of the main branch (in our case,
the lower-frequency component), and the deviation of the observed minima
in $G_{xx}$ from those of the main branch is at most 0.15$\pi$,
which is smaller than the accuracy of our LL index analysis. 


\begin{flushleft} 
{\bf S6. Energy band lineups}
\end{flushleft} 

A heterostructure is formed when a semiconductor is grown on top of
another semiconductor. The lineups of the conduction and valence bands
at the interface are of fundamental importance for engineering
semiconductor devices \cite{AlferovNL,KroemerNL,deWalle2003}. Since the
dawn of semiconductor technology, numerous models and theories have been
developed for calculating the energy band offsets. One of the first
attempts was the Anderson's electron affinity rule \cite{Anderson1962},
which is based on the consideration of the energy balance for an
electron moving from the vacuum energy level to the first semiconductor
(gaining $\chi_{1}$), then to the second semiconductor (losing $\Delta
E_{c}$), and then to the vacuum level again (losing $\chi_{2}$). From
the energy conservation, one obtains 
\begin{equation}
\Delta E_{c} = \chi_{1} - \chi_{2}, 
\end{equation}
where $\Delta E_{c}$ is the conduction-band offset and $\chi_{1}$
($\chi_{2}$) is the electron affinity of the first (second)
semiconductor. The valence-band offset follows automatically as 
\begin{equation}
\Delta E_{v} = E_{g2} - E_{g1} - \Delta E_{c}, 
\end{equation}
where $E_{g1}$ ($E_{g2}$) is the energy gap of the first (second)
semiconductor. The electron affinity rule was successful in explaining
the band discontinuities in many semiconductor heterostructures, but
failed for some, which is due to the following limitations of this model
\cite{Kroemer1985}: First, it idealizes the surface of a semiconductor.
In real crystals, the surface undergoes a reconstruction in order to
reduce the surface energy. The rearrangement of atoms leads to the
formation of a surface dipole layer which affects the electron affinity
of a semiconductor. Surface defects and surface energy states also play
roles. As a result, experimentally measured electron affinities are not
very reliable. (This is actually the case with both SnTe and
Bi$_{2}$Te$_{3}$, for which very different results have been reported in
the literature \cite{affBT1,affBT2,affST1,affST2,affST3,affST4,affST5}.)
Second, the surface reconstruction and surface energy states at the
interface of two semiconductors are not necessarily the same as those on
their free surfaces, making the energy balance consideration to be more
complicated. Third, electron correlation effects also influence the
electron affinity and have to be taken into consideration.

Over the years, there were many attempts to improve the description of
the band offsets at interfaces. One of them is known as ``common anion
rule" \cite{McCaldin1976,FK1977,K1982} and deals with heterostructures
which consist of semiconductors having a common anion element. This
approach is based on the assumption that the valence band is mostly
built from the atomic wave functions of anions, while the conduction
band is mostly built from the atomic wave functions of cations.
Therefore, the valence band of materials with the same anion element
should be similar, implying that their valence-band offset in the
heterostructure will be smaller than the conduction-band offset. This
might seem appropriate to the SnTe/Bi$_{2}$Te$_{3}$ interface; however,
the reality can be more complicated, especially in topological
insulators where the $s$- and $p$-orbital characters of the wave
functions forming the conduction and valence bands are interchanged in
comparison with ordinary materials. A good example demonstrating such a
band inversion is the (Sn$_{1-x}$Pb$_{x}$)Te system, in which a
topological phase transition has been demonstrated in ARPES experiments
\cite{Tanaka2012}.

Another approach to improve the simple electron affinity rule is the
effective dipole model \cite{Tersoff1986,HT1986,RuanChing1987}, which
includes the effects of the dipole charge formation due to a local
difference in the atomic (and electronic) structures at the interface in
comparison with the bulk structures of constituent semiconductors.

\begin{figure}
\begin{center}
\includegraphics[height=3.2cm]{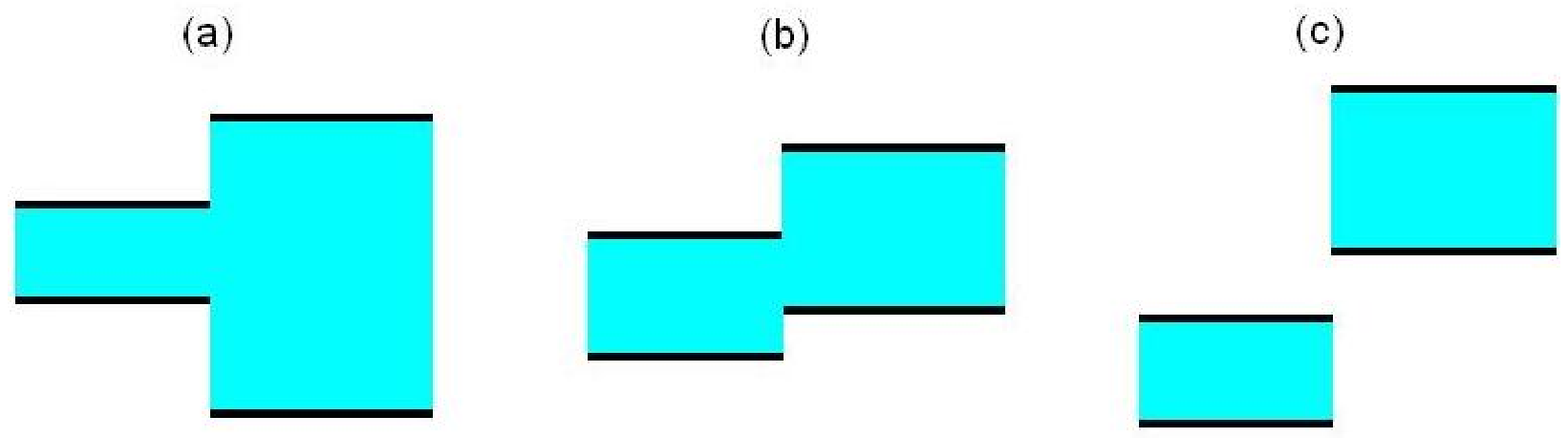}
\caption{
Schematic picture of three possible types of band alignments:
({\bf a}) straddling, ({\bf b}) staggered, and ({\bf c}) broken-gap band lineups.
} 
\end{center}
\end{figure}

Generally, there are only three possible band lineups \cite{KroemerNL}
which are shown in Fig. S3. The straddling lineup with conduction- and
valence-band offsets of opposite sign is the most common one. 
The staggered lineup has conduction- and valence-band offsets of the same 
sign. The broken-gap lineup is an extreme case of the staggered one, 
in which the bottom of the conduction band of one semiconductor goes below 
the top of the valence band of another semiconductor. This is the most exotic
lineup and it is realized in at least one nearly-lattice-matched
heterostructure, InAs/GaSb \cite{Sakaki1977}. Another example,
where the broken-gap lineup has been invoked to explain the coexistence
of $n$- and $p$-type carriers, is the case of quantum-well structures
and superlattices made of IV-VI semiconductors (such as PbTe/SnTe) for
thermoelectric applications \cite{affST2,affST3,affST4}. 


\begin{flushleft} 
{\bf S7. The broken-gap lineup}
\end{flushleft} 

To elucidate which of the three possible band lineups is realized in our
p-SnTe/n-Bi$_{2}$Te$_{3}$ heterojunction, measurements of the $I-V$
characteristics with current flowing through the interface are useful.
For both straddling and staggered lineups, an insulating barrier layer
will be formed at the interface as a result of the $p$-$n$ junction,
while for the broken-gap lineup, the system behaves as a semimetal
without any obstacles to the current at the interface. We therefore
prepared samples for perpendicular $I-V$ measurements in the following
way: First, rectangular-shaped islands were etched out from the
SnTe/n-Bi$_{2}$Te$_{3}$ film. Second, one half of each island was slowly
etched in the HCl:CH$_3$COOH:H$_2$O$_2$:H$_2$O solution in order to
remove the SnTe layer and a part of the Bi$_{2}$Te$_{3}$ buffer layer as
schematically shown in Fig. S4. Third, Pd contacts were deposited on top
of both the SnTe and Bi$_{2}$Te$_{3}$ layers (Fig. S4{\bf a}). The $I-V$
curves have been measured using a four-terminal dc-current method (Fig
S4{\bf b}). The resistivities of the
SnTe/Bi$_{2}$Te$_{3}$-heterojunction part and the
thinned-Bi$_{2}$Te$_{3}$-layer part of the islands have also been
measured separately using the van der Pauw method. For all samples, we
found linear $I-V$ characteristics with a negligible resistance of the
interface (the total resistance is mostly coming from the thinned
Bi$_{2}$Te$_{3}$ layer). Most likely, this result points to the
realization of the broken-gap lineup in the SnTe/Bi$_{2}$Te$_{3}$
heterojunction.

\begin{figure}
\begin{center}
\includegraphics[height=5.5cm]{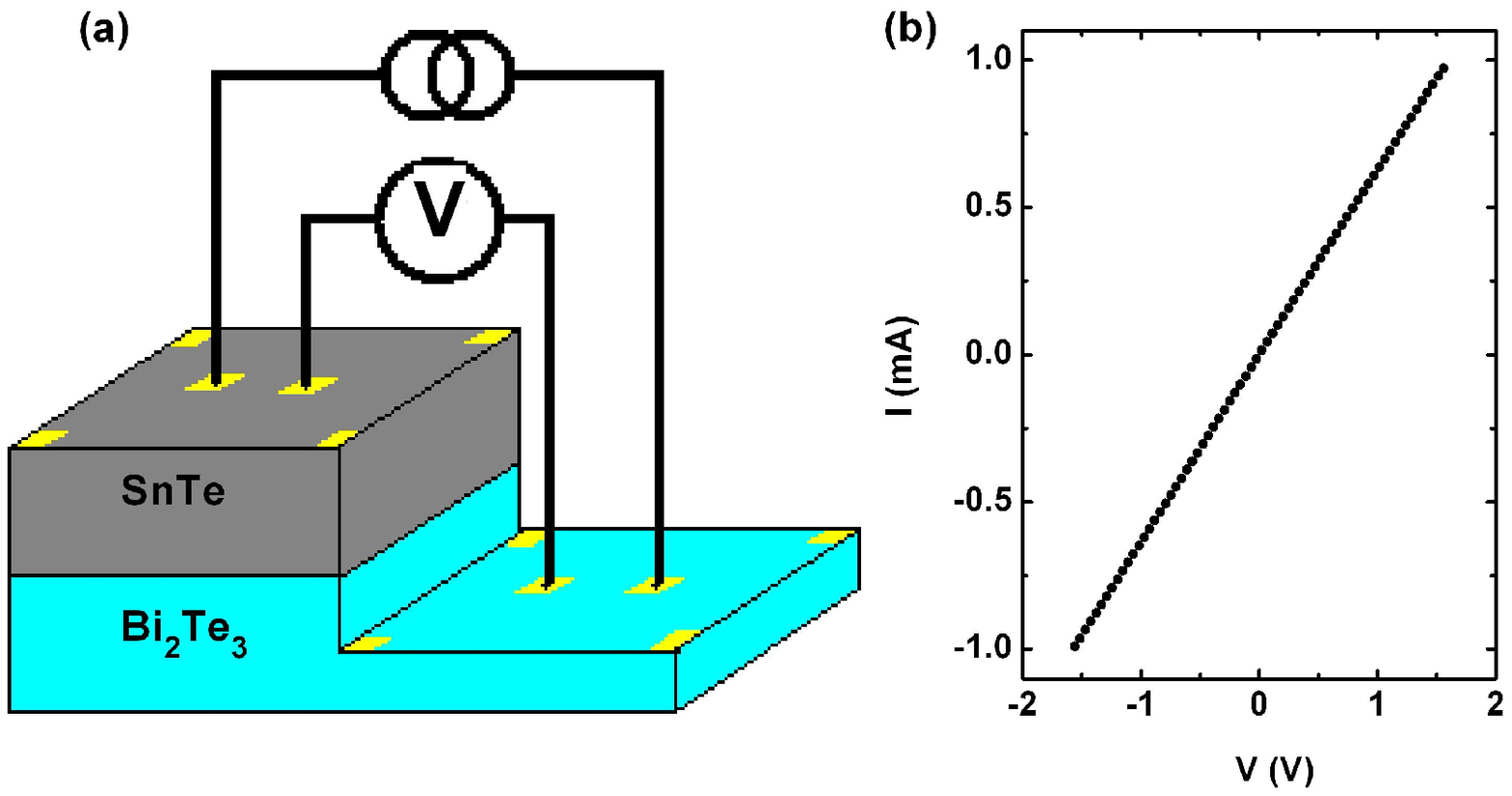}
\caption{
({\bf a}) Schematic picture of the sample and the experimental setup for the
$I-V$ measurements of the SnTe/Bi$_{2}$Te$_{3}$ interface. ({\bf b}) The $I-V$
characteristics of a typical SnTe/Bi$_{2}$Te$_{3}$ junction. } 
\end{center}
\end{figure}


\begin{flushleft} 
{\bf S8. Te-termination vs Sn-termination}
\end{flushleft} 

\begin{figure}[t]
\begin{center}
\includegraphics[width=8.5cm]{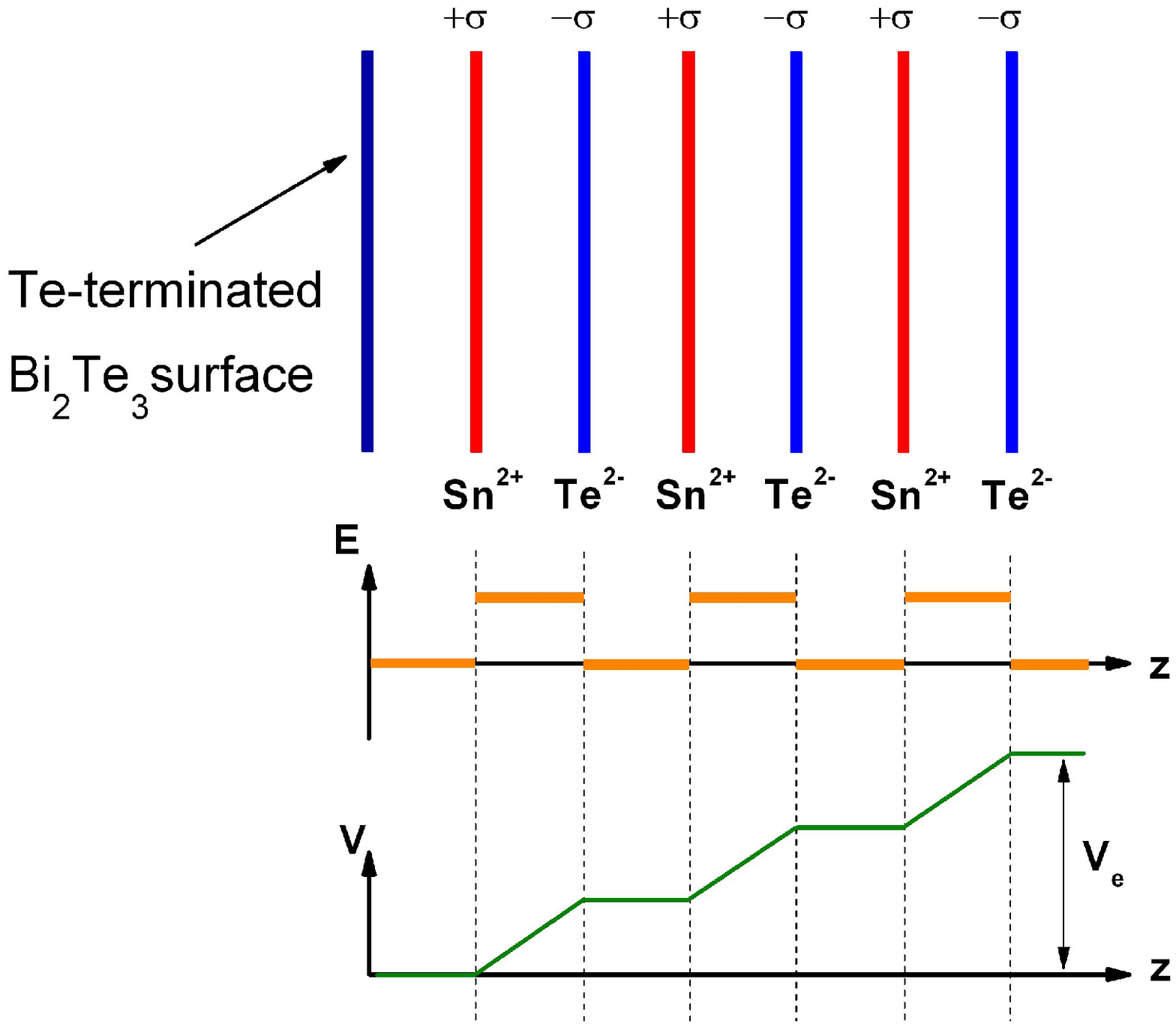}
\caption{
The polar catastrophe. A schematic charge distribution in idealized
ionic SnTe along the [111] direction is shown together with the
resulting variation of the electric fled $E$ and the electrostatic
potential $V$ along the [111] direction denoted as $z$. Note the
monotonic increase in $V$ with increasing number of atomic planes. } 
\end{center}
\end{figure}

SnTe is a material with partially ionic bonding. For films grown in
the [111] direction, the stacking sequence of atomic planes is
Sn$^{2+}$-Te$^{2-}$-Sn$^{2+}$-Te$^{2-}$-$\,\cdot\cdot\cdot$. Each
uniformly charged atomic plane generates the electric field
\begin{equation}
E =\frac{\sigma}{2 \epsilon \epsilon_0}
\end{equation}
where $+\sigma$ ($-\sigma$) is the charge density on the Sn$^{2+}$
(Te$^{2-}$) plane, $\epsilon_{0}$ is the permittivity of vacuum, and
$\epsilon$ is the permittivity of SnTe. (Note that the charge density
$\sigma$ is expected to be small, because the bonding in SnTe is mostly
covalent and is only partially ionic, due to the close electronegativity
of Sn and Te.) Such a charge distribution brings about a finite electric
field between each pair of Sn$^{2+}$ and Te$^{2-}$ planes, and, hence, a
monotonic increase in the electrostatic energy as shown in Fig. S5. For
increasing number of atomic layers along the polar direction, the
electrostatic energy diverges. This situation is known as a polar
catastrophe and cannot be realized in real materials \cite{Weiss}. One
way to avoid the polar catastrophe is to partially compensate the charge
on the surfaces of SnTe with mobile carriers as shown in Figs. S6 and
S7.

In our system, the SnTe layer starts with a Sn$^{2+}$ atomic
plane, which should be compensated with a negative charge. This is
naturally realized due to easily available electrons in the $n$-type
Bi$_{2}$Te$_{3}$ buffer layer. If the free surface is terminated with
Te, the negative charge of the outermost Te$^{2-}$ plane must be
compensated by positive charge (Fig. S6). The charge compensation makes
$E$ to change between positive and negative values along $z$, and the
resulting profile of $V$ just oscillates between zero and a finite
positive value. Nevertheless, in this Te-terminated case, there remains
a finite electrostatic potential $V_{e}$ at the outer surface (Fig. S6).

On the other hand, if the free surface is terminated with Sn, the
positive charge of the outermost Sn$^{2+}$ plane must be compensated by
negative charge (Fig. S7). In this case of the charge-compensated
Sn-terminated surface, the remaining electrostatic potential can be
exactly zero (Fig. S7). Compared with a finite $V_{e}$ in the case of
Te-terminated surface (Fig S6), the Sn-termination results in lower
electrostatic energy and thus is energetically favored. This means that
our films are more likely terminated with a Sn$^{2+}$ atomic plane.

We note that our transport data points to the existence of a small
density of high-mobility $n$-type carriers in the heterostructure, which
is only possible when the outer surface has a downward band bending to
cause the topological surface state to be doped with $n$-type carriers.
Such a band bending is actually expected for the Sn-terminated surface
which is compensated with negative charge, but the Te-terminated surface
will have an upward band-bending due to its compensation with positive
charge. Therefore, the conclusion about the termination based on the
electrostatic energy argument is supported by the transport results.

\begin{figure}
\begin{center}
\includegraphics[width=8.5cm]{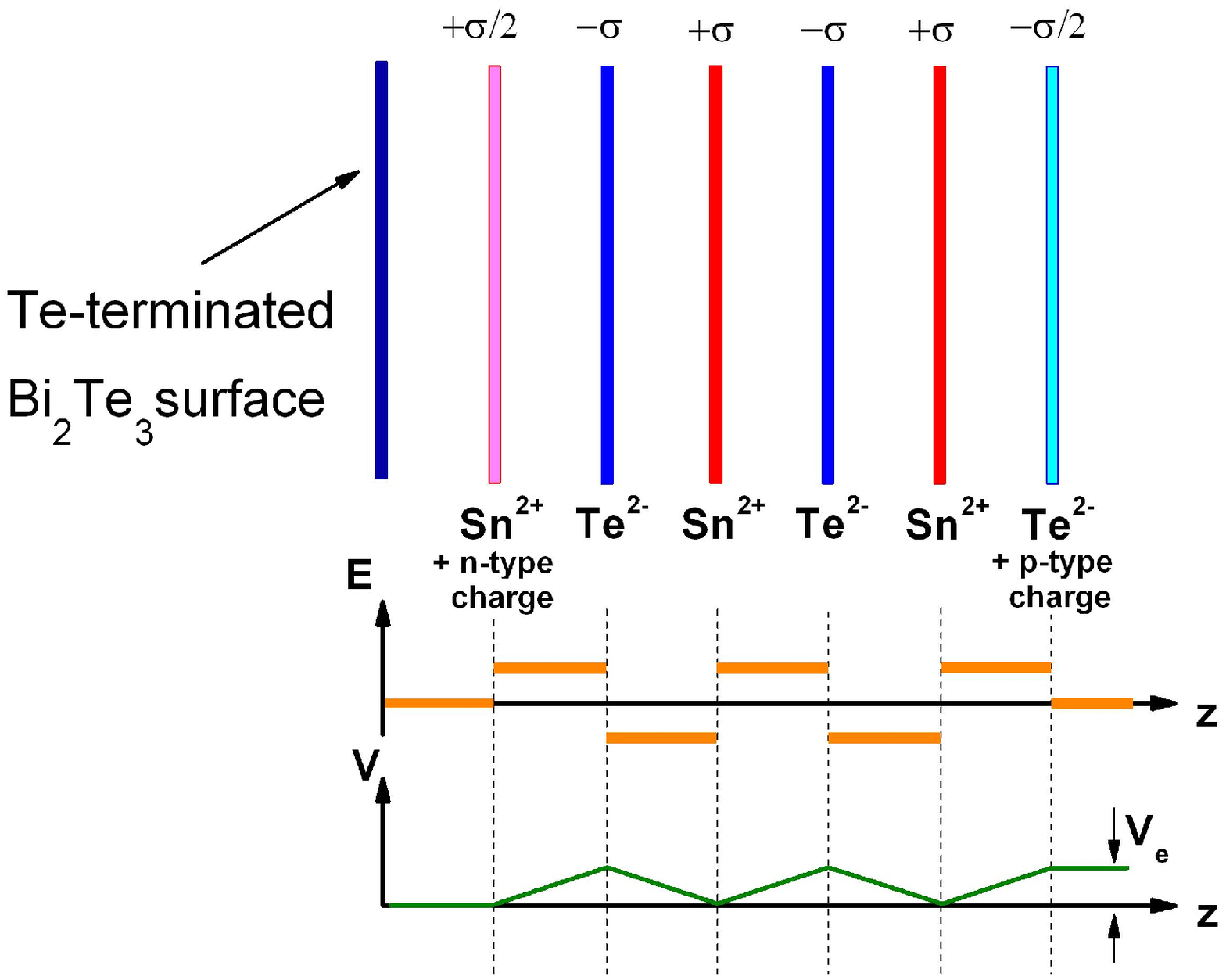}
\caption{
The situation for Te-terminated SnTe film with charge compensation.
A finite electrostatic potential $V_{e}$ remains at the outer surface.
} 
\end{center}
\end{figure}

\begin{figure}
\begin{center}
\includegraphics[width=8.5cm]{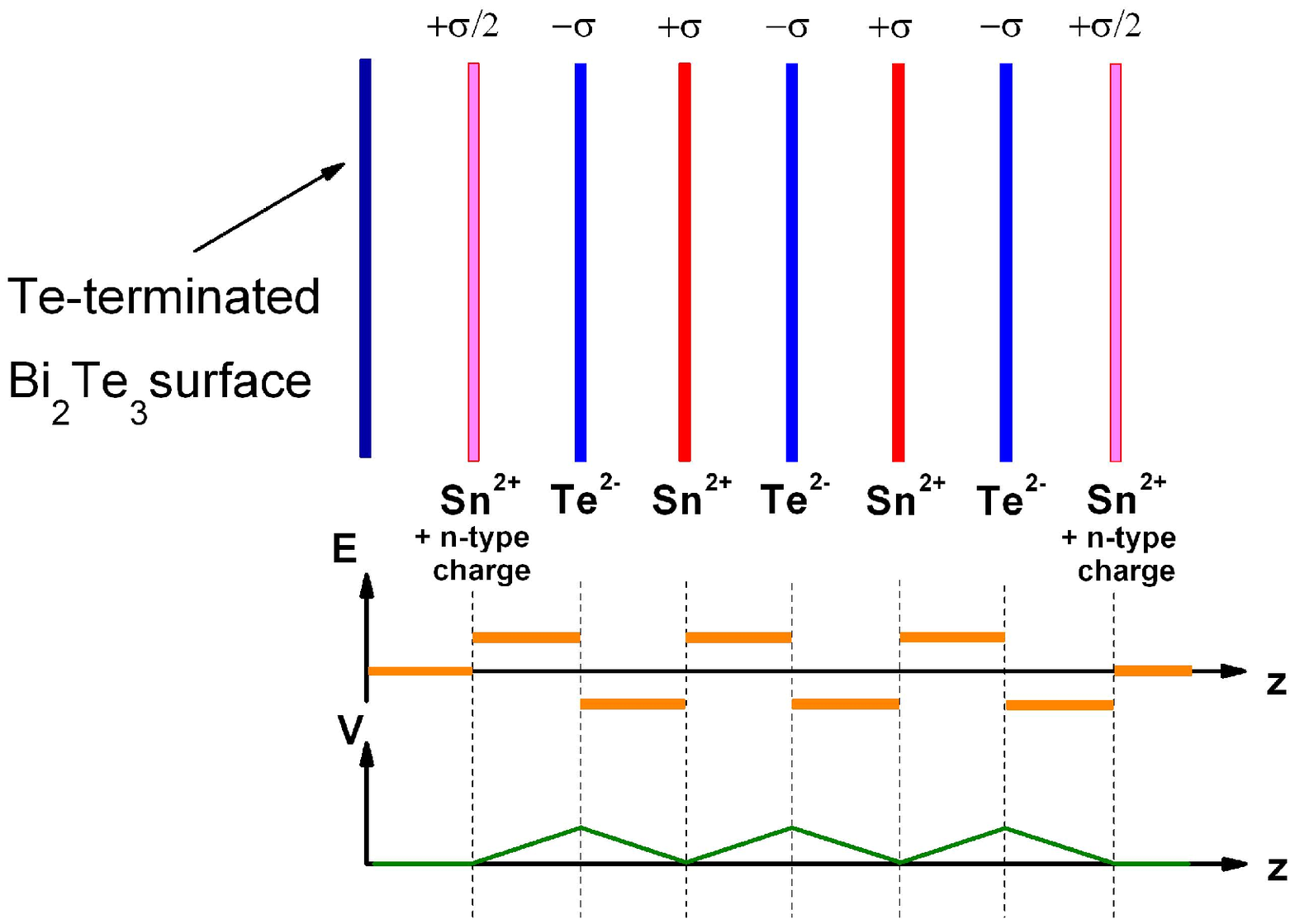}
\caption{
The situation for Sn-terminated SnTe film with charge compensation.
Remaining electrostatic potential $V_{e}$ is zero.
} 
\end{center}
\end{figure}

\end{document}